# Efficient and Effective Spam Filtering and Re-ranking for Large Web Datasets

**Gordon V. Cormack** · **Mark D. Smucker** ·
**Charles L. A. Clarke**

**Abstract** The TREC 2009 web ad hoc and relevance feedback tasks used a new document collection, the ClueWeb09 dataset, which was crawled from the general Web in early 2009. This dataset contains 1 billion web pages, a substantial fraction of which are spam — pages designed to deceive search engines so as to deliver an unwanted payload. We examine the effect of spam on the results of the TREC 2009 web ad hoc and relevance feedback tasks, which used the ClueWeb09 dataset. We show that a simple content-based classifier with minimal training is efficient enough to rank the "spamminess" of every page in the dataset using a standard personal computer in 48 hours, and effective enough to yield significant and substantive improvements in the fixed-cutoff precision (estP10) as well as rank measures (estR-Precision, StatMAP, MAP) of nearly all submitted runs. Moreover, using a set of "honeypot" queries the labeling of training data may be reduced to an entirely automatic process. The results of classical information retrieval methods are particularly enhanced by filtering — from among the worst to among the best.

## 1 Introduction

It is well-known that a vast number of web pages are created for the purpose of surreptitiously causing a search engine to deliver an unwelcome payload [10]. Broadly speaking, this purpose is effected by two mechanisms: self promotion and mutual promotion. A common form of self-promotion is word stuffing, in which gratuitous (often invisible to the reader) keywords are inserted to improve the retrieved rank of the page. A common form of mutual promotion is the link farm, in which a large number plausible-looking pages reference one another so as to improve a topic-independent quality score such as page rank. Often, the content of both kinds of spam pages is mechanically generated or plagiarized.

We are concerned with measuring and mitigating the effect of spam pages on retrieval effectiveness, as illustrated by two TREC 2009[1] tasks that use the new

---

University of Waterloo, Canada

[1] `trec.nist.gov`

ClueWeb09 dataset[2]. The first of these tasks is the ad hoc task of the Web Track [4]. The second task is the Relevance Feedback Track task, which evaluated relevance feedback techniques through a two-phase experimental protocol. Relevance judgments were provided to the participants after Phase 1, for use in Phase 2.

Definitions of Web spam typically focus on those pages that contain deceptive or harmful content. The authors of such spam pages may attempt to subvert the ranking algorithm of a search engine by presenting a false impression of relevance. However, "defining Web spam is not as straightforward as it might seem" [10], and it is often unclear whether a page is genuinely malicious or merely of low quality. We avoid considerations of author intent by employing a broad definition of Web spam that encompasses low-quality "junk" pages, those which are unlikely to be judged relevant to any query that might reasonably retrieve them.

The ClueWeb09 dataset was crawled from the general Web in early 2009, and contains roughly 1 billion pages written in a number of languages. The TREC tasks were concerned only with the English subset of about 500 million pages. Furthermore, several sub-tasks used a "Category B" subset containing about 50 million pages. The full set of 1 billion pages was dubbed "Category A." To our knowledge, all TREC participants submitting Category A runs used at most the 500 million page, English subset of ClueWeb09. The relevance feedback track specifically limited Category A runs to the English subset.

The literature lacks quantitative studies of the impact of spam and spam filtering on retrieval effectiveness. The AIRWeb Web Spam Challenge series[3] has measured the effectiveness of various methods at identifying spam hosts, but not the overall contribution of these methods to retrieval effectiveness. The Web Spam Challenge and other studies use two datasets prepared by Yahoo for the purpose: WEBSPAM-UK2006 and WEBSPAM-UK2007[4]. Each of these datasets consists of a crawl of a portion of the .uk web space, with spam or non-spam labels for a sample of a few thousand of the hosts represented in each crawl. To our knowledge, the only study of retrieval effectiveness using the WEBSPAM corpora [14] shows that users prefer spam-filtered results to unfiltered ones, but offers no relevance-based measurement of the impact of spam. The same authors investigate the properties that a collection should have to evaluate the effectiveness of "spam nullification" [15].

Previous TREC Web IR evaluation efforts have used corpora largely devoid of spam. For example, the TREC Terabyte track used a collection of government web pages [2]. Spam has been identified as an issue in the Blog Track [20], but its impact has not been systematically studied. The use of the ClueWeb09 dataset places the spam issue front and center at TREC for the first time. At least three participants [17, 18,11] used spam filters of some sort; one from a commercial search provider. Other participants noted the impact of spam on their efforts, particularly for Category A tasks [13,16,21].

Our objectives in undertaking this work were twofold: 1) to develop a practical method of labeling every page in ClueWeb09 as spam or not, and 2) to quantify the quality of the labeling by its impact on the effectiveness of contemporary IR methods. Our results are:

---

[2] boston.lti.cs.cmu.edu/Data/clueweb09

[3] webspam.lip6.fr/wiki/pmwiki.php

[4] barcelona.research.yahoo.net/webspam/datasets

- Several complete sets of spam labels, available for download without restriction. Each label is a percentile score which may be used in combination with a threshold to classify a page as "spam" or "not spam", or may be used to rank the page with respect to others by "spamminess."
- A general process for labeling large web datasets, which requires minimal computation, and minimal training.
  - A variant of the process is unsupervised, in that it uses automatically labeled training examples, with no human adjudication.
  - A variant uses training examples from a four-year-old, dissimilar, and much smaller collection.
  - A variant uses only 2.5 hours of human adjudication to label representative training examples.
  - A variant combines all of the above to yield a superior meta-ranking.
- Measurements that show a significant and substantive positive impact on precision at fixed cutoff, when the labels are used to remove spammy documents from all runs officially submitted by participants to the TREC Web adhoc and relevance feedback task.
- A method to automatically reorder, rather than simply to filter, the runs.
- Measurments that use 50-fold cross-validation to show a significant and substantive positive impact or reordering at all cutoff levels, and on rank-based summary measures.

Our measurements represent the first systematic study of spam in a dataset of the the magnitude of ClueWeb09, and the first quantitative results of the impact of spam filtering on IR effectiveness. Over and above the particular methods and measurements, our results serve as a baseline and benchmark for further investigation. New sets of labels may be compared to ours by the impact they have on effectiveness. Different methods of harnessing the labels — such as using them as a feature in learning to rank — may be compared to ours.

## 2 Context

Given the general consensus in the literature that larger collections yield higher precision [12], we expected the precision at rank 10 (P@10) scores for the TREC 2009 web ad hoc task to be high (P@10> 0.5), especially for the Category A dataset. Contrary to our expectations, the Category A results were poor (P@10 for all Cat A submitted runs: $\mu = 0.25, \sigma = 0.11, \max = 0.41$). The Category B results were better (P@10 for all Cat B submitted runs: $\mu = 0.38, \sigma = 0.07, \max = 0.56$), but still short of our expectation.

The authors represent two groups, X [7] and Y [24], that participated in TREC 2009, employing distinct retrieval methods. In the course of developing a novel method, group X composed a set of 67 pilot queries (Figure 1) and, for each, informally adjudicated the top few results from Category A. The results were terrible, the vast majority being spam. At the time, we took this to be a shortcoming of our method and reworked it using pseudo-relevance feedback from Wikipedia to yield higher quality results, which we characterized as "not terrible."

Group Y achieved satisfactory results (P@10=0.52) in Phase 1 of the relevance feedback (RF) task, which mandated Category B. In the final results, group X achieved strong performance on the Category A ad hoc task (P@10=0.38), while group Y did not (P@10=0.16). These results were surprising, as group Y's submission used exactly the same search engine and parameters as for relevance feedback — the only difference was the use of Category A instead of Category B.

**Table 1** Pilot Queries composed prior to TREC 2009.

| | | | |
|---|---|---|---|
| star wars | money spinal tap | apple | fish |
| star wars sdi | spinal tap | apple records | fishing |
| star wars luke | spinal tap procedure | apple computer | go fish |
| sdi | spinal tap lyrics | macintosh | fish episodes |
| selective disseminatinon | jaguar | macintosh apple | whip |
| spock | jaguar xj | apple macintosh | whip egg |
| spock kirk | jaguar cat | dead poets | whip crop |
| spock benjamin | jaguar fender | vacuum | whip topping |
| obama | fender | vacuum cleaner | party whip |
| obama japan | fender bender | high vacuum | whip it |
| barack obama | fender gibson | vacuum | bull whip |
| capital | gates | stream | WHIP 1350 AM |
| capital city | gates fences | stream process | whip flagellate |
| capital assets | gates steve | stream creek | chain whip |
| money | windows | stream education | The Whip |
| money pink floyd | windows doors | honda stream | whip antenna |
| money beatles | windows os | | WHIP walks hits |
| | | | inning pitched |

**Table 2** Group X and Group Y estimates of spam prevalence in top-ranked documents. Both estimates show that spam is highly prevalent in both Category A and Category B documents; more so in Category A.

| | Group X | Group Y |
|---|---|---|
| Category A | 612/756 = 0.81 (0.78 - 0.84) | 295/461 = 0.63 (0.59 - 0.68) |
| Category B | 47/74 = 0.64 (0.52 - 0.74) | 120/263 = 0.46 (0.39 - 0.52) |

A plausible explanation for these observations is that the TREC submissions were, in general, adversely affected by spam, and that the Category A collection has a higher proportion of spam than Category B. To validate this explanation, we first sought to quantify our observation that the top-ranked pages returned by our methods were dominated by spam. We then sought to find an automatic method to abate spam, and to evaluate the impact of that method on retrieval effectiveness.

To quantify the amount of spam returned, we constructed a web-based evaluation interface (Figure 1) and used it to adjudicate a number of the top-ranked pages returned during our preliminary investigation. Group X adjudicated 756 pages in total, selected at random with replacement from the top ten results for each of the 67 pilot topics. Group Y adjudicated 461 pages, selected at random with replacement from the top ten results for their Category A and Category B relevance feedback runs. These efforts consumed 2 hrs. 20 mins. and 1 hr. 20 mins. respectively. The results, shown with 95% confidence intervals in Table 2, indicate a high proportion of spam for the results of both groups in both categories. It follows that this high proportion of spam must have a substantial adverse effect on precision. As anticipated, the proportion of spam is higher in the Category A results.

**Fig. 1** User interface for adjudicating spamminess of CluWeb09 pages.

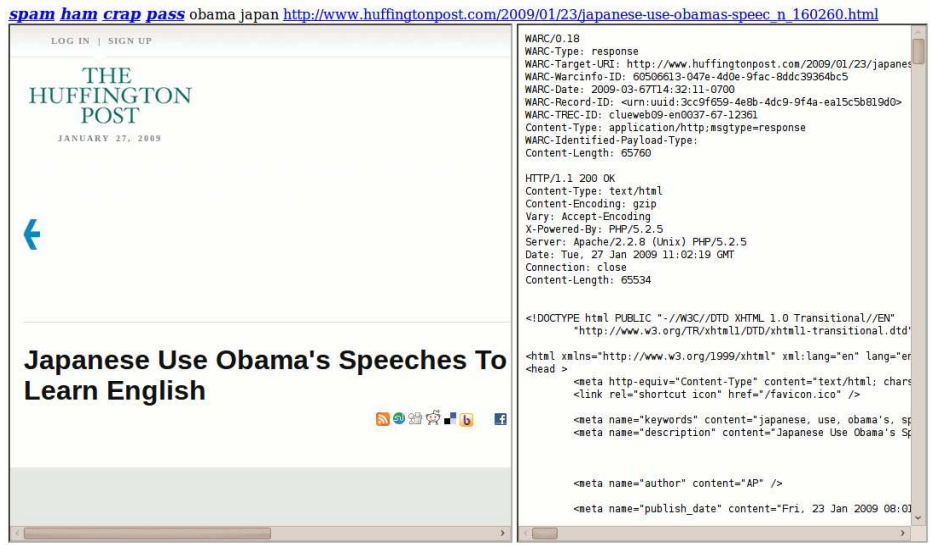

## 3 Evaluation Measures

The Group X and Group Y examples are entirely independent from each other, as they are derived from different topics and different retrieval methods, and assessed by different individuals. Furthermore, the Group X examples are independent of TREC, as the topics and pages were determined beforehand. For this reason, it is appropriate to use the Group X examples for training and tuning, and the Group Y examples for evaluation.

But evaluation on the Group Y examples answers only the narrow question, "How well can a spam filter identify spam in the pages returned by one particular method?" We are interested in the broader questions, "How well can a spam filter identify non-relevant pages?" and "What is the impact of removing or reranking these pages on retrieval effectiveness in general?"

For each of these questions, we require a suitable evaluation measure. For the first two — how well a filter identifies spam and nonrelevant documents — we use AUC, the area under the receiver operating characteristic curve. AUC — a threshold-independent measure of classifier effectiveness — has been used as the primary measure in previous web and email spam filtering evaluation efforts [6]. Our choice of evaluation measure for the third question — how well a filter improves retrieval effectiveness — was constrained by the sparsity of relevance assessments for TREC 2009. For the Category A ad hoc and relevance feedback tasks, the top-ranked 12 documents for each submitted run were assessed, as well as a stratified random sample of the rest. Precision at cutoff 10 (P@10) was reported as the primary retrieval effectiveness measure. These reported results serve as the baseline for our evaluation of filter impact.

Once documents are eliminated due to spam filtering, it is no longer the case that there are assessments for the top-ranked 12 documents, as lower ranked ones (which may not be assessed) rise in the ranking. It is therefore necessary to estimate P@10 for the filtered results, in order to compare them to the baseline. Furthermore, the Category B baseline must be estimated, as only a sample of the documents submitted for Category B submissions was assessed. We considered four methods of estimating P@10:

- *unjudged-nrel*. Unadjudicated documents considered to be nonrelevant. This method underestimates P@10.
- *unjudged-elided*. Unadjudicated documents are elided, and P@10 is computed on the top-ranked 10 documents that are adjudicated [23]. This method may underestimate P@10 because lower ranked documents are evaluated in place of higher ranked ones. Or it may overestimate P@10 because the elided documents are less likely to be relevant due to selection bias.
- *statPC10*. An intermediate calculation in the statAP method [3]:

$$statPC10 = \frac{statrel10}{10} \qquad (1)$$

  where $statrel10$ is a sample-based estimate of the number of relevant documents in the top-ranked 10. We determined experimentally that this method underestimates P@10 when the number of assessed documents is small. It yields an inexact estimate of $P@10$ even when the top 10 documents are fully adjudicated, and sometimes yields the absurd result $statPC10 > 1$.
- *estP10*. A sparse set-based estimate used for the TREC Legal Track [26]:

$$estP10 = \frac{estrel10}{\max(estrel10 + estnrel10, 1)} \qquad (2)$$

  where,

$$estrel10 = \min(statrel10, 10 - nrel10) \qquad (3)$$

  and

$$estnrel10 = \min(statnrel10, 10 - rel10). \qquad (4)$$

  Both $statrel10$ and $statnrel10$ are sample-based estimates of the number of relevant and non-relevant documents in the top-ranked 10, and $rel10$ and $nrel10$ are exact counts of the number of assessed relevant and non-relevant documents in the top-ranked 10. The $estP10$ measure yields more stable results than $statPC10$ and has the property that $estP10 = P@10$ when the top 10 documents are fully adjudicated. Moreover, it is nearly symmetric and therefore less likely to be biased: When none of the 10 documents is judged, $estP10 = 0$; otherwise, $estP10 = 1 - \overline{estP10}$ where $\overline{estP10}$ is calculated by complementing all judgements.

The results reported here use estP10 as the primary retrieval effectiveness measure, computed using the TREC Legal Track evaluation software `l07_eval` version 2.0[5]. The other three measures produce similar results and lead to the same conclusions. We briefly compare estP10 with the other measures in Section 5.

The estP10 measure estimates the effectiveness of retrieval for one specific task: identifying ten likely relevant documents. While this view of effectiveness is not unreasonable for web search, we are also concerned with effectiveness at cutoff values other than ten, and more generally *rank measures* that summarize effectiveness over

---

many cutoff values. The most commonly used rank measure is *mean average precision* (MAP) which is the mean of *average precision* (AP) over all topics:

$$AP = \frac{1}{R} \sum_k P@k \cdot rel(k) \,, \tag{5}$$

where $rel(k) = 1$ if the $k^{th}$-ranked document in the run is relevant; 0 if it is not. $R$ is the total number of relevant documents in the collection.

Unfortunately, $R$ is unknown for the TREC 09 tasks. Furthermore, $rel(k)$ is unknown for most $k$, in particular most $k > 12$. Methods to estimate AP with incomplete knowledge of $rel(k)$ have proven to be unreliable for the TREC 09 tasks [private correspondence, Web Track coordinators].

A more straightforward rank effectiveness measure is *R-precision*, (RP) which is simply

$$RP = P@R \,. \tag{6}$$

While R-precision depends on $R$, $R$ is not a direct factor in the formula, so estimation errors have much lower impact. Furthermore, $estRP$ is easily computed:

$$estRP = estP@estR \,. \tag{7}$$

Regardless of whether AP or RP or some other rank effectiveness measure is used, if our reranking improves $estPk$ for all values of $k$, it follows that the rank measure is improved. Therefore, as our primary evaluation of the effectiveness of spam reranking, we compute $estPk$ for representative values of $k$. As our best effort to quantify the magnitude of the effect, we present $estRP$ as well. Furthermore, we evaluate StatMAP, MAP (unjudged nonrelevant), and MAP (unjudged elided).

## 4 Spam Filter Design

Our principal criteria in choosing a spam filter design were efficiency and effectiveness. By efficiency, we mean the end-to-end time and resource consumption (both human and computer) to label the corpus. By effectiveness, we mean the ability to identify spam (and hence nonrelevant) pages among those retrieved documents and thus to improve precision by deleting it.

Although the literature is dominated by graph-based methods for web spam filtering and static ranking [1,22], content-based email spam filters were found to work as well as graph-based methods in the 2007 Web Spam Challenge [5]. Furthermore, these filters are very fast, being able to classify thousands of documents per second. Our implementation required about 48 hours elapsed time to decompress, decode and score the 500M English ClueWeb09 pages on a standard PC with Intel dual core E7400 CPU.

We used three different sets of training examples to create three filters, each of which was used to label the entire corpus; in addition, we created an ensemble filter using a naive Bayes metaclassifier to combine the results:

– *UK2006.* The WEBSPAM-UK2006 corpus, used for the AIRWeb Web Spam Challenge and other studies, contains spam and nonspam labels for 8238 hosts. For each spam host and each nonspam host, we selected the first page in the corpus whose size was at least 5K bytes. This approach tends to select an important page near the root of the host's web space. Our training set consisted of 767 spam pages and 7474 nonspam pages — one for each spam host and one for each nonspam host. Our aim in using this set of training examples was to investigate the efficacy of *transfer learning* from an older, smaller, less representative corpus.

**Table 3** Top 40 queries from the 1000 queries used for the *Britney* training examples.

| | | | |
|---|---|---|---|
| 1. britney spears | 11. pamela anderson | 21. ipod | 31. carmen electra |
| 2. youtube | 12. angelina jolie | 22. coach | 32. wikipedia |
| 3. facebook | 13. lindsay lohan | 23. american idol | 33. runescape |
| 4. wwe | 14. jennifer hallett | 24. nfl | 34. pokemon |
| 5. barack obama | 15. hi-5 | 25. jessica alba | 35. hannah montana |
| 6. kim kardashian | 16. clay aiken | 26. miley cyrus | 36. john mccain |
| 7. myspace | 17. iphone | 27. limewire | 37. online dictionary |
| 8. sarah palin | 18. xbox 360 | 28. dragonball | 38. stock market |
| 9. naruto | 19. wii | 29. megan fox | 39. club penguin |
| 10. paris hilton | 20. psp | 30. nba | 40. webkinz |

– *Britney.* Our second set of training examples was essentially generated automatically, requiring no manual labeling of spam pages. We asked ourselves, "If we were spammers, where would we find keywords to put into our pages to attract the most people?" We started looking for lists of popular searches and found an excellent source at a search engine optimization (SEO) site[6]. This particular SEO site collects the daily published "popular search queries" from the major web search engines, retailers, and social tagging sites. We used their collected Google Trends, Yahoo! Buzz, Ask, Lycos, and Ebay Pulse queries. We downcased all queries and took the top 1000 for the year 2008, the period immediately before the ClueWeb09 corpus was crawled. The most popular query was "britney spears" and hence the name of this training data. Table 3 shows other query examples. We used the `#combine` operator in Indri [25] to perform naive query likelihood retrievals from Category A with these 1000 queries. We used the same index and retrieval parameters as [24]. For each query, we took the top ten documents and summarily labeled them as spam, with no human adjudication. We fetched the Open Directory Project archive[7] and intersected its links with the URIs found in ClueWeb09. From this intersection, we selected 10,000 examples which we summarily labeled as nonspam, with no human adjudication. Our rationale for using this set of training examples was derived from the observation that our naive methods retrieved almost all spam, using the queries that we composed prior to TREC. We surmised that popular queries would be targeted by spammers, and thus yield an even higher proportion of spam — high enough that any non-spam would be within the noise tolerance limits of our spam filter. In effect, the SEO queries acted as a "honeypot" to attract spam.

– *Group X.* We used the 756 documents adjudicated by Group X as training examples (table 2, column 1). Recall that these documents were selected without knowledge of the TREC topics or relevance assessments. Our objective was to determine how well a cursory labeling of messages selected from the actual corpus would work.

– *Fusion.* The scores yielded by the three filters were interpreted as log-odds estimates and averaged, in effect yielding a naive Bayes combination of the three scores. This approach is known to be effective for both email and web spam filtering [5, 19].

We expected all the filters to identify spam better than chance, but had no prediction as to which set of training examples would work best. In particular, we did not know how well training on the UK2006 examples would transfer to ClueWeb09 due to the differences in the times of the crawls, the hosts represented, and the representativeness of the host-based labels. We did not know how well "pages retrieved by a naive method in response to a popular query" would act as proxies for spam, or how overfitted to the particular queries the results would be. Similarly, we did not know how well ODP pages would act as proxies for non-spam. We did not know if the Group X examples were sufficiently numerous, representative, or carefully labeled to yield a good classifier. We did have reason to think that the fusion filter might outperform all the rest, consistent with previously reported results.

---

[6] `www.seomoz.org/popular-searches/index/2008-`$mm$-$dd$

[7] `rdf.dmoz.org`

## 4.1 Filter operation

A linear classifier was trained using on-line gradient-descent logistic regression in a single pass over the training examples [9]. The classifier was then applied to the English portion of the ClueWeb09 dataset end-to-end, yielding a spamminess score for each successive page $p$. Owing to the use of logistic regression for training, the spamminess score may be interpreted as a log-odds estimate:

$$score(p) \approx \log \frac{\Pr[p\,is\,spam]}{\Pr[p\,is\,nonspam]}\,. \tag{8}$$

However, this estimate is likely to be biased by the mismatch between training and test examples. Nonetheless, a larger score indicates a higher likelihood of spam, and the sum of independent classifier scores is, modulo an additive constant, a naive Bayes estimate of the combined log-odds.

For the purpose of comparing effectiveness, we convert each score to a percentile rank over the 503,903,810 English pages:

$$percentile(p) = \left\lfloor 100 \frac{|p'|score(p') \geq score(p)|}{503,903,810} \right\rfloor\,. \tag{9}$$

That is, the set of pages with $percentile(p) < t$ represents the spammiest $t\%$ of the corpus. In the results below, we measure effectiveness for $t \in [0, 100]$, where $t = 0$ filters nothing and $t = 100$ filters everything.

## 4.2 Filter implementation

The implementation of the classifier and update rule are shown in Figure 2. Apart from file I/O and other straightforward housekeeping code, these figures contain the full implementation of the filter.[8] The function `train()` should be called on each training example. After training, the function `spamminess()` returns a log-odds estimate of the probability that the page is spam.

Each page, including WARC and HTTP headers, was treated as flat text. No tokenization, parsing, or link analysis was done. Pages exceeding 35,000 bytes in length

---

[8] Figure 2 is Copyright © 2010 Gordon V. Cormack. This code is free software: you can redistribute it and/or modify it under the terms of the GNU General Public License as published by the Free Software Foundation, either version 3 of the License, or (at your option) any later version.

**Fig. 2** C implementation of the filter. The function `spamminess` is the soft linear classifier used for spam filtering. The function `train` implements the online logistic regression gradient descent training function.

```
#define P 1000081
#define PREF 35000
float w[P];
unsigned dun[P], cookie;
float spamminess(unsigned char *page, int n){
    unsigned i, b, h;
    cookie++;
    if (n > PREF) n = PREF;
    float score=0;
    b = (page[0]<<16) | (page[1]<<8) | page[2];
    for (i=3;i<n;i++){
        b = (b<<8) | page[i];
        h = b %
        if (dun[h] == cookie) continue;
        dun[h] = cookie;
        score += w[h];
    }
    return score;
}

#define delta 0.002
train(unsigned char *page, int n, int IsSpam){
    unsigned i, b, h;
    if (n > PREF) n = PREF;
    float p=1/(1+exp(-spamminess(page,n)));
    cookie++;
    b = (page[0]<<16) | (page[1]<<8) | page[2];
    for (i=3;i<n;i++){
        b = (b<<8) | page[i];
        h = b %
        if (dun[h] == cookie) continue;
        dun[h] = cookie;
        w[h] += (IsSpam-p) * delta;
    }
}
```

were arbitrarily truncated to this length. Overlapping byte 4-grams were used as features. That is, if the page consisted of "pq xyzzy" the features would be simply "pq x", "q xy", " xyz ", "xyzz", and "yzzy". Each feature was represented as a binary quantity indicating its presence or absence in the page. Term and document frequencies were not used. Finally, the feature space was reduced from $4 \times 10^9$ to $10^6$ dimensions using hashing and ignoring collisions. This brutally simple approach to feature engineering was used for one of the best filters in the TREC 2007 email spam filtering task [8], giving us reason to think it would work here.

Given a page $p$ represented by a feature vector $X_p$ a linear classifier computes

$$score(p) = \beta \cdot X_p \qquad (10)$$

where the weight vector $\beta$ is inferred from training examples. For the particular case of on-line gradient-descent logistic regression, the inference method is quite simple. $\beta$ is initialized to $\overline{0}$, and for each training document $p$ in arbitrary order, the following

**Table 4** ROC Area (AUC) and 95% confidence intervals for three base filters, plus the naive Bayes fusion of the three. The "Relevance" column reflects the ability of the filter to remove non-relevant documents.

| | Category A | | Category B | |
|---|---|---|---|---|
| | Spam | Relevance | Spam | Relevance |
| UK2006 | 0.94 (0.91 - 0.95) | 0.85 (0.81 - 0.88) | 0.90 (0.86 - 0.93) | 0.79 (0.73 - 0.84) |
| Britney | 0.88 (0.85 - 0.91) | 0.87 (0.84 - 0.90) | 0.80 (0.75 - 0.84) | 0.78 (0.74 - 0.83) |
| GroupX | 0.92 (0.89 - 0.94) | 0.91 (0.88 - 0.94) | 0.89 (0.84 - 0.93) | 0.87 (0.83 - 0.91) |
| Fusion | 0.95 (0.93 - 0.97) | 0.93 (0.90 - 0.95) | 0.94 (0.91 - 0.96) | 0.86 (0.83 - 0.92) |

update rule is applied:

$$\beta \leftarrow \beta + \delta X_p \left( isspam(p) - \frac{1}{1 + e^{-score(p)}} \right) , \text{ where} \tag{11}$$

$$isspam(p) = \begin{cases} 1 & p \, is \, spam \\ 0 & p \, is \, nonspam \end{cases} . \tag{12}$$

We fixed the learning rate parameter $\delta = 0.002$ based on prior experience with email spam and other datasets.

## 5 Filter Results

We first consider how well the four filters identify spam, and how well they rank for static relevance. We then consider the impact on the TREC 2009 web ad hoc submissions on average, and the impact on the individual submissions. Finally, we consider the impact on the TREC 2009 relevance feedback submissions.

Figure 3 shows the effectiveness of the UK2006 filter at identifying spam in the examples labeled by Group Y. The top and middle panels show the fraction of spam pages identified, and the fraction of nonspam pages identified, as a function of the percentile threshold. For a good filter, the lines should be far apart, indicating that a great deal of spam can be eliminated while losing little nonspam. The two panels indicate that the filter is effective, but do little to quantify how far apart the lines are. The bottom panel plots the first curve as a function of the second — it is a receiver operating characteristic (ROC) curve. The area under the curve (AUC) – a number between 0 and 1 — indicates the effectiveness of the filter. The results (0.94 for Category A and 0.90 for Category B) are surprisingly good, comparable to the best reported for the 2007 AIRWeb Challenge. AUC results for this and the other three filters are shown, with 95% confidence intervals, in Table 4. These results indicate that all filters are strong performers, with UK2006 and GroupX perhaps slightly better at identifying spam than Britney. The fusion filter, as predicted, is better still.

Figure 4 shows the filter's effectiveness at identifying nonrelevant documents, as opposed to spam. To measure nonrelevance, we use the same documents discovered by Group Y, but the official TREC relevance assessments instead of Group Y's spam labels. We see that the curves are well separated and the AUC scores are only slightly lower than those for spam identification. Table 4 summarizes the AUC results for both spam and static relevance.

We would expect high correlation between documents identified as spam and documents assessed to be non-relevant, but were surprised nonetheless by how well the filter worked for this purpose. These results suggest that spam is a strong predictor — perhaps the principal predictor – of nonrelevance in the results returned by a search engine.

**Fig. 3** Effect of filtering on elimination of spam and nonspam. The top and centre columns show the fraction of spam and nonspam eliminated as a function of the fraction of the corpus that is labeled "spam." The bottom panel shows the corresponding ROC curves.

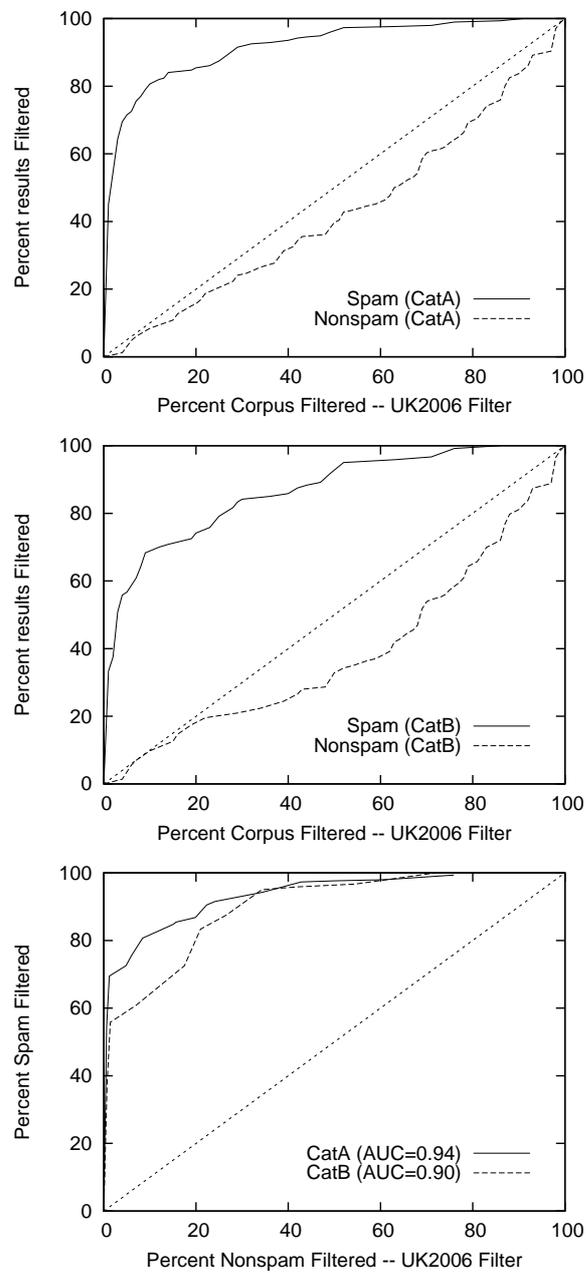

**Fig. 4** Effect of filtering on elimination of nonrelevant and relevant pages. The top and centre columns show the fraction of nonrelevant and relevant pages eliminated as a function of the fraction of the corpus that is labeled "spam." The bottom panel shows the corresponding ROC curves.

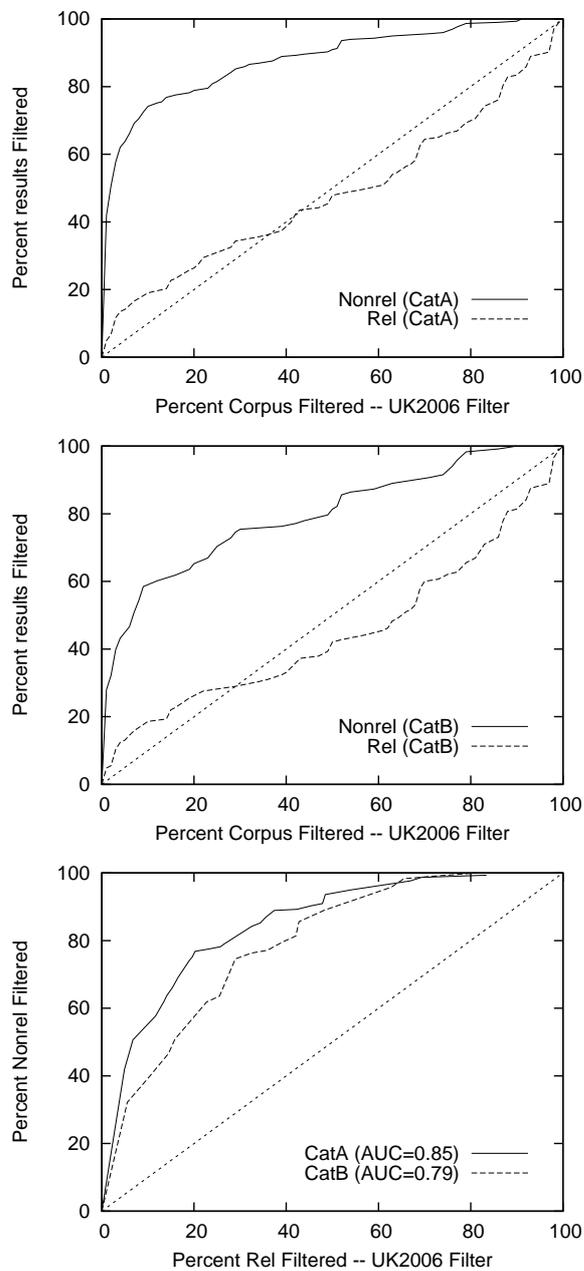

**Fig. 5** Effect of spam filtering on the average effectiveness of all Web Track ad hoc submissions, Category A and Category B. Effectiveness is shown as precision at 10 documents returned (P@10) as a function of the fraction of the corpus that is labeled "spam".

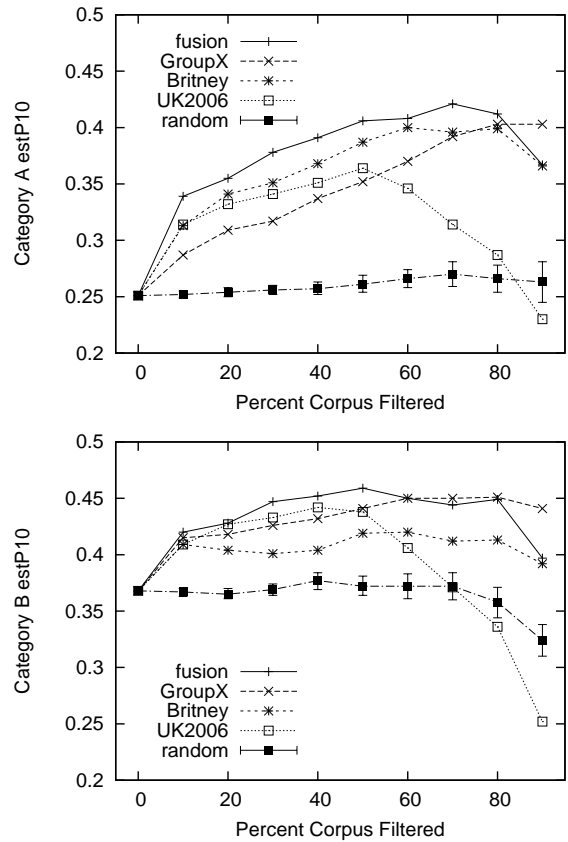

To evaluate the impact of filtering on retrieval effectiveness, we acquired from the TREC organizers all submissions for the TREC 2009 web ad hoc and relevance feedback tasks. We applied the four filters — and also a random control — for threshold settings of $t \in 0, 10, 20, 30, 40, 50, 60, 70, 80, 90$. The random control simply labeled $t\%$ of the corpus at random to be spam. Our prediction was that for an effective filter, $estP10$ should increase with $t$ and eventually fall off. For the random control, $estP10$ should either remain flat or fall off slightly, assuming the submissions obey the probability ranking principle.

Figure 5 shows $estP10$, averaged over all official TREC submissions, as a function of $t$ for each of the filters. All (except the control) rise substantially and then fall off as predicted. The control appears to rise insubstantially, and then fall off. It is entirely possible that the rise is due to chance, or that the probability ranking is compromised by the presence of very highly ranked spam. 95% confidence intervals are given for the control, but omitted for the other filters as their superiority is overwhelmingly significant ($p \ll 0.001$).

**Fig. 6** Effect of spam filtering on the effectiveness of individual Web Track ad hoc submissions, Category A and Category B. The top scatterplot shows P@10 with 70% of the corpus labeled "spam" by the fusion method; the bottom scatterplot shows P@10 with 50% of the corpus labeled "spam" by the fusion method. The correlations with 95% confidence intervals for the top and bottom plots respectively are 0.09 (-0.24–0.40) and 0.60 (0.34–0.79).

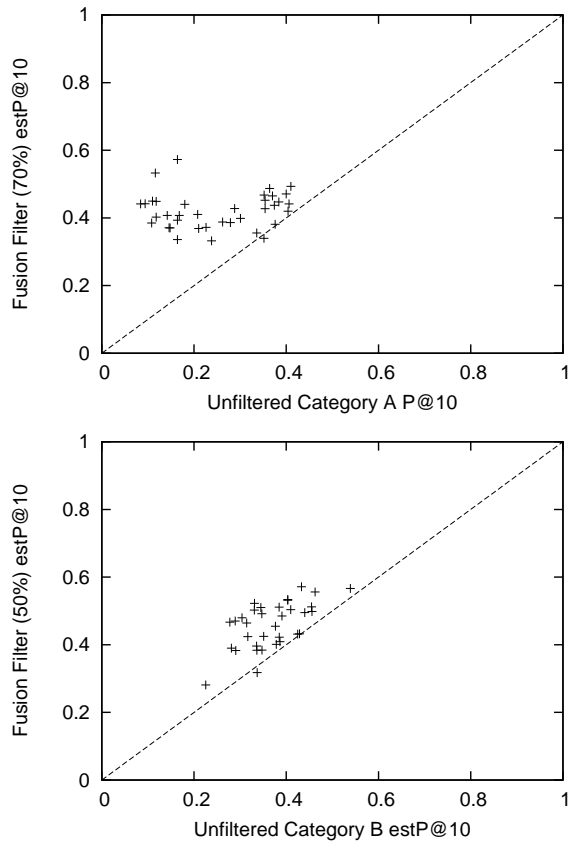

All filters behave as predicted. $estP10$ increases to $t = 50$, at which point the UK2006 filter starts to fall off. Beyond $t = 50$, the other filters continue to improve for Category A, and plateau for Category B. As expected, the fusion filter is superior to the rest, reaching peak effectiveness at $t = 70$ for Category A and $t = 50$ for Category B. The fusion filter with these threshold settings is used to illustrate the impact on individual TREC submissions.

The UK2006 filter is trained on documents from a different corpus, which exclude corpus-specific information like WARC and HTTP headers. The strong performance of this filter — notwithstanding the falloff at high thresholds — is evidence that spam, and not some artifact of the data itself, is responsible for the results presented here.

Figure 6 shows a scatterplot comparing unfiltered with filtered $estP10$ results. Nearly every submission is improved by filtering. The left panel (Category A) is partic-ularly remarkable as it shows no significant correlation between filtered and unfiltered results for particular runs (95% confidence interval: $-0.24 - 0.40$). That is, the effect of spam filtering overwhelms any other differences among the submissions. Tables 5 and 6 respectively report the results of the fusion filter for the individual Category A and Category B ad hoc runs.

**Table 5** Effect of spam filtering on the TREC web ad hoc submissions, Category A. Shown are the estP10 results for 3 threshold settings of the fusion filter: 0% (no filtering), 50%, and 70%.

| Run ID | No filter | estP10 50% filter | 70% filter |
|---|---|---|---|
| MS1 | 0.3540 | 0.4042 | 0.4273 |
| MS2 | 0.4060 | 0.4241 | 0.4414 |
| MSRAAF | 0.3540 | 0.4195 | 0.4522 |
| MSRAC | 0.4000 | 0.4368 | 0.4710 |
| MSRANORM | 0.3700 | 0.4286 | 0.4652 |
| Sab9wtBase | 0.2260 | 0.3147 | 0.3720 |
| Sab9wtBf1 | 0.2880 | 0.3572 | 0.4277 |
| Sab9wtBf2 | 0.2620 | 0.3562 | 0.3878 |
| THUIR09An | 0.3740 | 0.4191 | 0.4372 |
| THUIR09LuTA | 0.2100 | 0.3270 | 0.3688 |
| THUIR09TxAn | 0.3640 | 0.4699 | 0.4873 |
| UMHOObm25GS | 0.1420 | 0.3918 | 0.4073 |
| UMHOObm25IF | 0.1640 | 0.3084 | 0.3359 |
| UMHOOqlGS | 0.1180 | 0.3832 | 0.4024 |
| UMHOOqlIF | 0.1080 | 0.3994 | 0.3849 |
| WatSdmrm3 | 0.1180 | 0.3916 | 0.4490 |
| WatSdmrm3we | 0.1640 | 0.5913 | 0.5725 |
| WatSql | 0.0840 | 0.4207 | 0.4416 |
| muadanchor | 0.3519 | 0.4313 | 0.4678 |
| muadibm5 | 0.2788 | 0.3737 | 0.3864 |
| muadimp | 0.3006 | 0.3760 | 0.3988 |
| pkuLink | 0.1160 | 0.4786 | 0.5330 |
| pkuSewmTp | 0.1480 | 0.3771 | 0.3704 |
| pkuStruct | 0.1460 | 0.3753 | 0.3710 |
| twCSrs9N | 0.2080 | 0.3821 | 0.4104 |
| twCSrsR | 0.1800 | 0.4428 | 0.4403 |
| twJ48rsU | 0.2380 | 0.3296 | 0.3323 |
| uogTrdphP | 0.1680 | 0.4369 | 0.4075 |
| uvaee | 0.1100 | 0.4444 | 0.4499 |
| uvamrf | 0.0940 | 0.4113 | 0.4420 |
| uvamrftop | 0.4100 | 0.4903 | 0.4935 |
| watprf | 0.3360 | 0.3342 | 0.3553 |
| watrrfw | 0.3760 | 0.3774 | 0.3813 |
| watwp | 0.3516 | 0.3476 | 0.3396 |
| yhooumd09BFM | 0.1640 | 0.3984 | 0.3933 |
| yhooumd09BGC | 0.3840 | 0.5049 | 0.4472 |
| yhooumd09BGM | 0.4040 | 0.4819 | 0.4198 |

Figure 7 illustrates the dramatic impact of spam filtering on our simple query likelihood method. In Category A, our submission is improved from the worst unfiltered result to better than the best unfiltered result. In Category B, the same method (which was not an official submission to TREC) sees a less dramatic but substantial improvement.

Figure 8 shows the effect of filtering on the relevance feedback runs. The baseline results are stronger, but still improved substantially by filtering.

**Table 6** Effect of spam filtering on the TREC web ad hoc submissions, Category B. Shown are the estP10 results for 3 threshold settings of the fusion filter: 0% (no filtering), 50%, and 70%.

| | estP10 | | |
| Run ID | No filter | 50% filter | 70% filter |
| --- | --- | --- | --- |
| ICTNETADRun3 | 0.4332 | 0.5715 | 0.5438 |
| ICTNETADRun4 | 0.4402 | 0.4951 | 0.5150 |
| ICTNETADRun5 | 0.3907 | 0.4851 | 0.5076 |
| IE09 | 0.3864 | 0.4092 | 0.4238 |
| NeuLMWeb300 | 0.4557 | 0.4984 | 0.4587 |
| NeuLMWeb600 | 0.4096 | 0.5038 | 0.4714 |
| NeuLMWebBase | 0.3040 | 0.4796 | 0.4604 |
| RmitLm | 0.3307 | 0.5023 | 0.4526 |
| RmitOkapi | 0.3509 | 0.4249 | 0.3949 |
| SIEL09 | 0.3784 | 0.4012 | 0.4158 |
| UCDSIFTinter | 0.4286 | 0.4322 | 0.4478 |
| UCDSIFTprob | 0.3849 | 0.4222 | 0.4106 |
| UCDSIFTslide | 0.4248 | 0.4315 | 0.4459 |
| UDWAxBL | 0.3166 | 0.4238 | 0.3940 |
| UDWAxQE | 0.3369 | 0.3175 | 0.2884 |
| UDWAxQEWeb | 0.4625 | 0.5562 | 0.5157 |
| UMHOObm25B | 0.3846 | 0.5114 | 0.4691 |
| UMHOOqlB | 0.3449 | 0.5098 | 0.4506 |
| UMHOOsd | 0.4033 | 0.5315 | 0.5103 |
| UMHOOsdp | 0.4033 | 0.5335 | 0.5123 |
| UamsAw7an3 | 0.3766 | 0.4546 | 0.4521 |
| UamsAwebQE10 | 0.3142 | 0.4643 | 0.4324 |
| arsc09web | 0.2254 | 0.2813 | 0.2073 |
| irra1a | 0.2776 | 0.4666 | 0.4320 |
| irra2a | 0.2905 | 0.3827 | 0.4123 |
| irra3a | 0.2893 | 0.4699 | 0.4396 |
| scutrun1 | 0.3362 | 0.3839 | 0.4539 |
| scutrun2 | 0.3474 | 0.3841 | 0.4283 |
| scutrun3 | 0.3358 | 0.3960 | 0.4466 |
| udelIndDMRM | 0.3312 | 0.5223 | 0.4846 |
| udelIndDRPR | 0.2811 | 0.3899 | 0.3653 |
| udelIndDRSP | 0.3469 | 0.4917 | 0.4616 |
| uogTrdphA | 0.4548 | 0.5119 | 0.4581 |
| uogTrdphCEwP | 0.5389 | 0.5664 | 0.5496 |

Figure 9 recasts the superior and inferior curves from Figure 5 in terms of the other three measures. The overall effect is the same for all measures: filtering substantially improves P@10 over baseline for a wide range of threshold settings.

## 6 Reranking Method

In the experiments reported in the previous section, we used the score returned by our classifier in the crudest possible way: as a brick wall filter that effectively eliminates some fraction of the corpus from consideration. Here, we consider instead the problem of using the spam scores to reorder the ranked list of documents returned by a search engine.

In reranking, we do not eliminate documents with high scores; instead we move them lower in the ranking. Presumably, documents with extreme scores should be moved more than others, but by how much? Our approach is to use supervised learning to compute the best new ranking, given the original ranking and the spam percentile scores.

**Fig. 7** Effect of filtering on naive query likelihood language model runs.

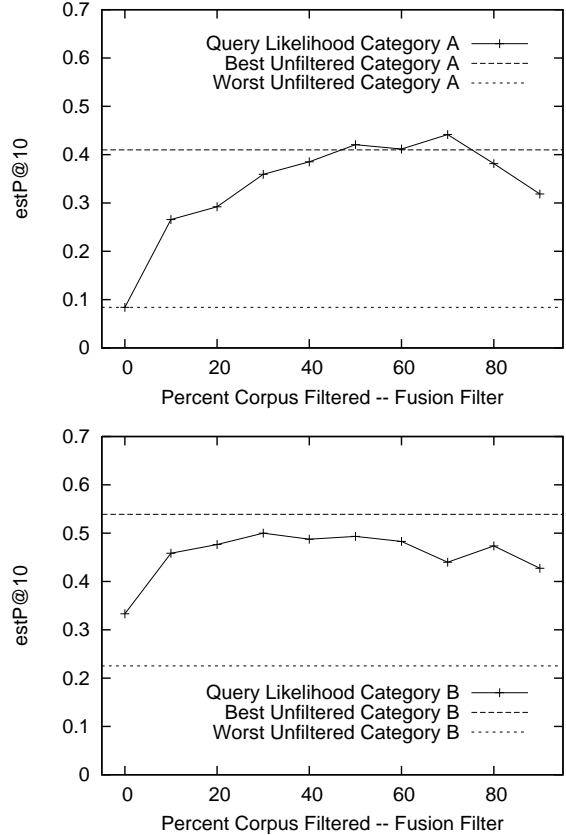

Supervised learning requires training examples. In a real-world deployment, the training examples would be constructed by adjudicating the results of historical queries presented to the same search engine. For our experiments, we have no historical results — only those from the 50 topics used at TREC 2009, which have no particular historical order. We therefore use 50-fold cross-validation, using one topic at a time for evaluation, and considering the remaining 49 to be historical examples. The evaluation results from these 50 separate experiments – each reranking the results for one topic — are then averaged.

Our learning method, more properly learning-to-rank method, consists of exhaustive enumeration to compute, for all $k$, the threshold $t_k$ that optimizes $estPk$:

$$t_k = \arg\max_t estPk \text{ where } t = t_k \,. \tag{13}$$

**Fig. 8** Effect of spam filtering on the average effectiveness of all relevance feedback task Phase 2 submissions, Category A and Category B. Effectiveness is shown as precision at 10 documents returned (P@10) as a function of the fraction of the corpus that is labeled "spam".

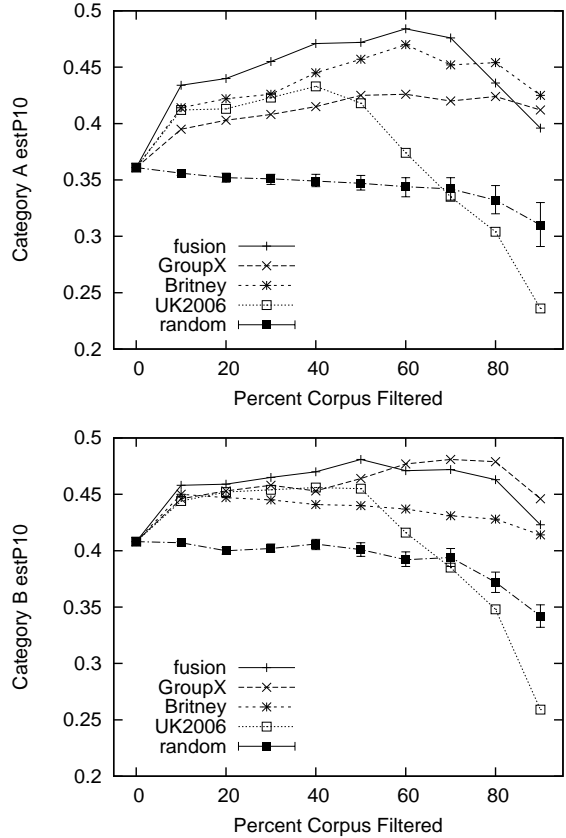

Then we proceed in a greedy fashion to build the new ranked list $r'$ from the original $r$:

$$r'[1] = r[\min\{i | score(r[i]) \geq t_k\}] \tag{14}$$

$$r'[i > 1] = r[\min i | score(r[i]) \geq t_k \text{ and } r[i] \notin r'[1, i-1]] \text{ , except} \tag{15}$$

when equation 15 is undefined, in which case

$$r'[i > 1] = r[i] \, . \tag{16}$$

The special case is occasioned by the fact that $t_k$ is not necessarily monotonic in $k$ due to noise in the training examples.

This reranking method was applied independently to each TREC ad hoc Category A submission, using our fusion filter's percentile scores. Table 7 shows $estP30$, $estP300$ and $estRP$ for each pair of original and reranked results, along with the average over all runs and the P-value for the difference. Figure 8 shows StatMAP, MAP (unjudged not relevant), and MAP (unjudged elided). All measures show a substantial improvement for nearly all runs. The notable exception is the run labeled *watwp* which, ironically, is the author's submission. This run consists entirely of Wikipedia documents, so it is not surprising that spam filtering does not improve it! *watprf* is the only other run that is not improved in terms of both *estRP* and MAP (unjudged elided), leading us to suggest that the other few cases are outliers, perhaps due to to the bias of incomplete relevance assessment.

**Fig. 9** Other P@10 estimates for Category A ad hoc runs, fusion filter.

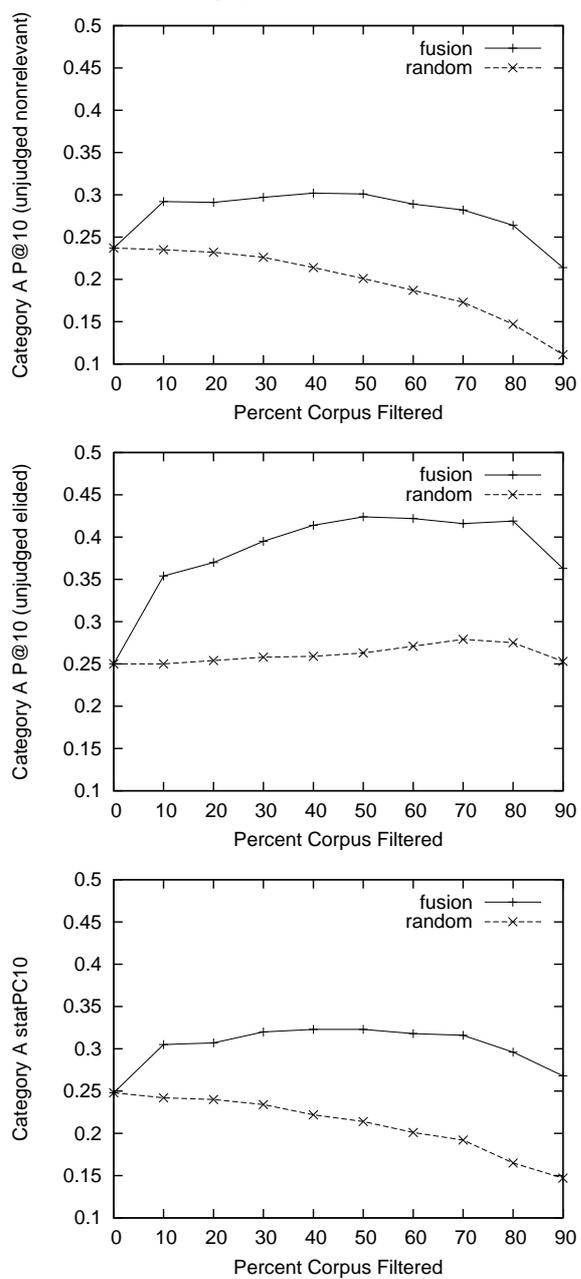

**Table 7** Effect of spam reranking on TREC ad hoc submissions, Category A. Representative cutoff depths and R-precision show improvement at all ranks.

| Run ID | estP30 orig. | estP30 rerank | estP300 orig | estP300 rerank | estRP orig | estRP rerank |
|---|---|---|---|---|---|---|
| MS1 | 0.3728 | **0.4029** | 0.3506 | **0.3841** | 0.2999 | **0.3282** |
| MS2 | **0.3921** | 0.3891 | 0.3702 | **0.3837** | 0.3479 | **0.3646** |
| MSRAAF | 0.3751 | **0.4271** | 0.3793 | **0.4280** | 0.3748 | **0.3944** |
| MSRAC | 0.3925 | **0.4397** | 0.3807 | **0.4284** | 0.3891 | **0.4074** |
| MSRANORM | 0.3949 | **0.4370** | 0.3812 | **0.4294** | 0.3751 | **0.3975** |
| Sab9wtBase | 0.2472 | **0.3248** | 0.3061 | **0.3119** | 0.2788 | **0.2880** |
| Sab9wtBf1 | 0.3031 | **0.3526** | 0.3219 | **0.3407** | 0.3113 | **0.3147** |
| Sab9wtBf2 | 0.2737 | **0.3105** | 0.3154 | **0.3226** | 0.2932 | **0.3010** |
| THUIR09An | 0.3485 | **0.3575** | 0.3008 | **0.3062** | 0.2858 | **0.2924** |
| THUIR09LuTA | 0.2321 | **0.3493** | 0.3123 | **0.3893** | 0.2965 | **0.3298** |
| THUIR09TxAn | 0.3356 | **0.4189** | 0.3411 | **0.4094** | 0.3188 | **0.3474** |
| UMHOObm25GS | 0.1653 | **0.4121** | 0.2586 | **0.3523** | 0.2540 | **0.3334** |
| UMHOObm25IF | 0.1640 | **0.3105** | 0.2329 | **0.3297** | 0.2201 | **0.2897** |
| UMHOOqlGS | 0.1359 | **0.4238** | 0.2402 | **0.3246** | 0.2380 | **0.3160** |
| UMHOOqlIF | 0.1326 | **0.4069** | 0.2396 | **0.3364** | 0.2373 | **0.3236** |
| WatSdmrm3 | 0.1224 | **0.4221** | 0.2361 | **0.3290** | 0.2431 | **0.3132** |
| WatSdmrm3we | 0.1905 | **0.5587** | 0.3062 | **0.4267** | 0.3090 | **0.3927** |
| WatSql | 0.1111 | **0.4229** | 0.2356 | **0.3188** | 0.2361 | **0.3117** |
| muadanchor | 0.3316 | **0.3909** | 0.3311 | **0.3737** | 0.3097 | **0.3367** |
| muadibm5 | 0.2881 | **0.3638** | 0.3376 | **0.3935** | 0.3539 | **0.3727** |
| muadimp | 0.2886 | **0.3716** | 0.3372 | **0.3930** | 0.3513 | **0.3694** |
| pkuLink | 0.1581 | **0.4766** | 0.2694 | **0.3887** | 0.2558 | **0.3353** |
| pkuSewmTp | 0.1600 | **0.3863** | 0.2558 | **0.3429** | 0.2473 | **0.3319** |
| pkuStruct | 0.1595 | **0.3894** | 0.2514 | **0.3408** | 0.2469 | **0.3320** |
| twCSrs9N | 0.2376 | **0.2818** | 0.2945 | **0.3719** | 0.2800 | **0.3189** |
| twCSrsR | 0.2059 | **0.4180** | 0.2962 | **0.3607** | 0.2807 | **0.3318** |
| tw.J48rsU | 0.2501 | **0.2877** | 0.2727 | **0.2994** | 0.2610 | **0.2802** |
| uogTrdphP | 0.2194 | **0.4561** | 0.2994 | **0.4146** | 0.2764 | **0.3606** |
| uvaee | 0.1400 | **0.4160** | 0.2709 | **0.3538** | 0.2808 | **0.3432** |
| uvamrf | 0.1178 | **0.4285** | 0.2564 | **0.3512** | 0.2493 | **0.3426** |
| uvamrftop | 0.3516 | **0.4468** | 0.3049 | **0.3668** | 0.3057 | **0.3493** |
| watprf | 0.3040 | **0.3081** | 0.2976 | **0.3033** | 0.3067 | **0.3047** |
| watrrfw | 0.3511 | **0.3513** | 0.2826 | **0.2854** | 0.2950 | **0.2989** |
| watwp | **0.3389** | 0.3303 | **0.2310** | 0.2310 | 0.2311 | **0.2332** |
| yhooumd09BFM | 0.1828 | **0.4018** | 0.2580 | **0.3666** | 0.2532 | **0.3299** |
| yhooumd09BGC | 0.3520 | **0.4804** | 0.3102 | **0.3851** | 0.2817 | **0.3311** |
| yhooumd09BGM | 0.3689 | **0.4594** | 0.3065 | **0.3811** | 0.2795 | **0.3284** |
| Average | 0.2567 | **0.3949** | 0.2965 | **0.3582** | 0.2880 | **0.3318** |
| P-value (1-tailed) | < 0.0001 | | < 0.0001 | | < 0.0001 | |

## 7 Discussion

While it is common knowledge that the purpose of web spam is to subvert the purpose of information retrieval methods, a relevance-based quantitative assessment of its impact — or of methods to mitigate its impact — has not previously been reported.

**Table 8** Effect of spam reranking on TREC ad hoc submissions, Category A. Representative cutoff depths and R-precision show improvement at all ranks.

| Run ID | StatMAP | | MAP (unj. nonrel.) | | MAP (unj. elided) | |
|---|---|---|---|---|---|---|
| | orig. | rerank | orig | rerank | orig | rerank |
| MS1 | 0.0419 | **0.0488** | 0.0597 | **0.0646** | 0.1346 | **0.1479** |
| MS2 | 0.0619 | **0.0655** | 0.0806 | **0.0811** | 0.1554 | **0.1609** |
| MSRAAF | 0.0484 | **0.0538** | 0.0754 | **0.0783** | 0.1489 | **0.1639** |
| MSRAC | 0.0534 | **0.0605** | 0.0795 | **0.0858** | 0.1550 | **0.1715** |
| MSRANORM | 0.0509 | **0.0572** | 0.0763 | **0.0813** | 0.1537 | **0.1695** |
| Sab9wtBase | 0.0225 | **0.0244** | 0.0338 | **0.0310** | 0.0783 | **0.0886** |
| Sab9wtBf1 | **0.0388** | 0.0382 | **0.0480** | 0.0436 | 0.1031 | **0.1088** |
| Sab9wtBf2 | 0.0274 | **0.0292** | **0.0403** | 0.0376 | 0.0866 | **0.0914** |
| THUIR09An | 0.0301 | **0.0312** | 0.0460 | **0.0477** | 0.0826 | **0.0869** |
| THUIR09LuTA | 0.0378 | **0.0438** | 0.0490 | **0.0524** | 0.1146 | **0.1385** |
| THUIR09TxAn | 0.0480 | **0.0562** | 0.0653 | **0.0717** | 0.1353 | **0.1586** |
| UMHOObm25GS | 0.0414 | **0.0651** | 0.0641 | **0.0920** | 0.1295 | **0.1866** |
| UMHOObm25IF | 0.0273 | **0.0354** | 0.0385 | **0.0496** | 0.0971 | **0.1322** |
| UMHOOqlGS | 0.0338 | **0.0600** | 0.0554 | **0.0900** | 0.1179 | **0.1791** |
| UMHOOqlIF | 0.0307 | **0.0552** | 0.0518 | **0.0856** | 0.1116 | **0.1754** |
| WatSdmrm3 | 0.0338 | **0.0547** | 0.0580 | **0.0839** | 0.1261 | **0.1786** |
| WatSdmrm3we | 0.0596 | **0.0885** | 0.0919 | **0.1284** | 0.1717 | **0.2498** |
| WatSql | 0.0310 | **0.0557** | 0.0517 | **0.0861** | 0.1159 | **0.1775** |
| muadanchor | 0.0148 | **0.0246** | 0.0225 | **0.0302** | 0.0553 | **0.0645** |
| muadibm5 | 0.0396 | **0.0553** | 0.0443 | **0.0596** | 0.1467 | **0.1702** |
| muadimp | 0.0394 | **0.0559** | 0.0444 | **0.0607** | 0.1476 | **0.1728** |
| pkuLink | 0.0319 | **0.0552** | 0.0429 | **0.0690** | 0.1000 | **0.1582** |
| pkuSewmTp | 0.0361 | **0.0582** | 0.0543 | **0.0773** | 0.1131 | **0.1667** |
| pkuStruct | 0.0376 | **0.0596** | 0.0566 | **0.0796** | 0.1154 | **0.1703** |
| twCSrs9N | 0.0168 | **0.0259** | 0.0271 | **0.0363** | 0.0861 | **0.1046** |
| twCSrsR | 0.0232 | **0.0390** | 0.0370 | **0.0510** | 0.1014 | **0.1316** |
| twJ48rsU | 0.0200 | **0.0228** | 0.0262 | **0.0286** | 0.0653 | **0.0720** |
| uogTrdphP | 0.0406 | **0.0623** | 0.0627 | **0.0878** | 0.1322 | **0.1905** |
| uvaee | 0.0475 | **0.0804** | 0.0653 | **0.0961** | 0.1371 | **0.2011** |
| uvamrf | 0.0355 | **0.0641** | 0.0579 | **0.0961** | 0.1201 | **0.1869** |
| uvamrftop | 0.0867 | **0.0974** | 0.1058 | **0.1198** | 0.1832 | **0.2187** |
| watprf | 0.0628 | **0.0633** | **0.0633** | 0.0602 | **0.1082** | 0.1087 |
| watrrfw | 0.0723 | **0.0738** | 0.0834 | **0.0834** | 0.1539 | **0.1593** |
| watwp | 0.0486 | **0.0496** | **0.0496** | 0.0491 | **0.0772** | 0.0770 |
| yhooumd09BFM | 0.0248 | **0.0394** | 0.0357 | **0.0544** | 0.1006 | **0.1495** |
| yhooumd09BGC | 0.0442 | **0.0552** | 0.0674 | **0.0773** | 0.1359 | **0.1667** |
| yhooumd09BGM | 0.0385 | **0.0478** | 0.0593 | **0.0672** | 0.1259 | **0.1518** |
| Average | 0.0400 | **0.0528** | 0.0560 | **0.0696** | 0.1195 | **0.1510** |
| P-value (1-tailed) | < 0.0001 | | < 0.0001 | | < 0.0001 | |

Measurements of the prevalence of spam in retrieved results and the ability of filters to identify this spam give some indication but do not tell the whole story. The bottom line is: How much does spam hurt, and how much is this hurt salved by spam filtering? For both questions, we offer a lower bound which is substantial. A simple on-line logistic regression filter dramatically improves the effectiveness of systems participating in the TREC 2009 web ad hoc and relevance feedback tasks, including those from major Web search providers, some of which employ their own spam filters. One may infer from the improvement that the impact of spam is similarly substantial. Unless, that is, the spam filter is learning some aspect of page quality apart from spamminess. We find this explanation unlikely, as the AUC scores indicate that the filters indeed identify spam.

In any event the distinction is moot: If the filters have serendipitously discovered some other aspect of static relevance, what is the harm?

At the time of writing, graph-based quality and spam metrics for ClueWeb09 were still in preparation by a third party [private communication, Carlos Castillo, Andras Benczur], and we did not have the means to anticipate the results of this major undertaking. When they do become available, they may be easily compared to and perhaps combined with ours to form a super meta-filter.

Several of the TREC 2009 submissions already incorporated some form of spam filtering. The *yhooumd00BGM* run, for example [18], used Yahoo's spam labels and a learning method to improve its P@10 score from 0.1420 to 0.4040. Our reranking improves it further to 0.4724. The authors of *uvamrftop [17]* also paid particular attention to spam; our method improves its result from 0.4100 to 0.4855. *twJ48rsU* [11]is similarly improved from 0.2380 to 0.2801.

For those who wish to experiment with any of these approaches, or simply to apply the filters as described here, we make the four sets of percentile scores available for download. Using a custom compressor/decompressor, each is about 350MB compressed and 16GB uncompressed[9].

**Acknowledgements** The authors thank Ellen Voorhees and Ian Soboroff at the National Institute of Standards and Technology (U.S.A) for providing access to the TREC data. Invaluable feedback on a draft was provided by Stephen Tomlinson, Ian Soboroff and Ellen Voorhees. This research was supported by grants from the Natural Sciences and Engineering Research Council (Canada) and from Amazon.

---

[9] `durum0.uwaterloo.ca/clueweb09spam`